\documentclass[a4paper]{article}

\usepackage{a4wide}
\usepackage[UKenglish]{babel}
\usepackage{graphicx}
	\graphicspath{{./}{figures/}}
\usepackage[colorlinks=true,linkcolor=blue,citecolor=red]{hyperref}
\usepackage{xcolor}
\usepackage{amsmath,amsthm,bbold,multirow,siunitx}
\usepackage[inline]{enumitem}

\theoremstyle{definition}\newtheorem{assumption}{Assumption}[section]
\theoremstyle{remark}\newtheorem{remark}[assumption]{Remark}

\newcommand{\E}{\mathbb{E}}
\newcommand{\R}{\mathbb{R}}

\allowdisplaybreaks

\begin{document}

\title{Reconstruction of traffic speed distributions from kinetic models with uncertainties}

\author{Michael Herty\thanks{RWTH Aachen University, Templergraben 55, 52056 Aachen, Germany} \and
		Andrea Tosin\thanks{Politecnico di Torino, Corso Duca degli Abruzzi 24, 10129 Torino, Italy} \and
		Giuseppe Visconti\thanks{RWTH Aachen University, Templergraben 55, 52056 Aachen, Germany} \and
		Mattia Zanella\thanks{University of Pavia, Via A. Ferrata 5, 27100 Pavia, Italy}}
\date{}

\maketitle

\begin{abstract}
In this work we investigate the ability of a kinetic approach for traffic dynamics to predict speed distributions obtained through rough data. The present approach adopts the formalism of uncertainty quantification, since reaction strengths are uncertain and linked to different types of driver behaviour or different classes of vehicles present in the flow. Therefore, the calibration of the expected speed distribution has to face the reconstruction of the distribution of the uncertainty. We adopt experimental microscopic measurements recorded on a German motorway, whose speed distribution shows a multimodal trend. The calibration is performed by extrapolating the uncertainty parameters of the kinetic distribution via a constrained optimisation approach. The results confirm the validity of the theoretical set-up.

\medskip

\end{abstract}

\section{Introduction}
Similarly to classic rarefied gas dynamics, kinetic modelling for traffic flow needs to define basic dynamics on microscopic entities. In particular, traffic ``particles'' are the vehicles, which modify their speed according to some binary interaction laws, whose definition may impact on the aggregate description and on the limit hydrodynamic trends, see~\cite{Guenter2002,Herty2010,Herty2018,Klar1997,puppo2017CMS,tosin2018MMS}. Furthermore, when describing behavioural phenomena the physics-inspired methods of kinetic theory needs to face new challenges, since interaction forces cannot be inferred from first principles and physical forces are replaced by empirical social forces. These new interactions are typically deduced heuristically with the aim to reproduce the qualitative behaviour of the system and are at best known with the aid of statistical methods. 

Once a sound kinetic model is available, its effectivity can be measured in terms of its ability to replicate and forecast system dynamics. Nevertheless, the uncertainty which is present at the level of particles may have a very strong effect at different scales. In addition, the most used methods coming from uncertainty quantification, such as generalised polynomial chaos or collocation methods, typically assume the knowledge of the uncertainty distribution in order to develop accurate solvers, see~\cite{Dimarco2017,Hu2017,Xiu2010}. Unfortunately, structural uncertainties in social systems may be highly non-standard and may change in time due to external influences.

In this chapter we aim at theoretical insights into the extrapolation of the statistical distribution of the uncertainty starting from data on traffic dynamics collected within the project~\cite{dataset}. In particular, we will try to calibrate the speed distribution predicted by a kinetic traffic model taking advantage of the knowledge of the empirical one obtained from real data. We will show how the microscopic uncertainty is naturally transferred to the observable quantities and, based on the measured mixed traffic conditions, we will propose an approach that catches the empirical speed distributions in several traffic regimes. We think that the promising results produced by the present approach may be useful both for the prediction and for the accurate reconstruction of real phenomena after sensitivity analysis.

This chapter has the following structure: in Section~\ref{sec:2} we introduce the theoretical set-up of the problem with emphasis on the role of the uncertainty in the modelling of microscopic interactions. We also introduce a Boltzmann-type kinetic approach for traffic dynamics and, in a suitable asymptotic regime, we compute the equilibrium speed distributions, which depend on the microscopic uncertainty. Moreover, we define the quantities of interest to be compared with the traffic data described in Section~\ref{sec:3}. Finally, in Section~\ref{sec:4} we perform the calibration of the equilibrium distributions in several traffic regimes through constrained optimisation techniques. 

\section{Kinetic modelling with uncertain interactions}
\label{sec:2}
Traditional microscopic traffic models are based on the assumption that the traffic stream is composed by homogeneous vehicles, whose reaction to speed changes is linked to the vehicle type. However, structural differences between vehicles are often observed in real traffic flows in terms, for example, of vehicle weight, engine efficiency and more or less aggressive driver behaviour. The traffic heterogeneity influences the deceleration/acceleration process of drivers in mixed traffic conditions, see~\cite{Munigety2016,Munigety2018}. Experimental evidence of this fact and of the relation with traffic safety issues has been recently presented in~\cite{WHO2018}. 

In the following we will show that, thanks to the theoretical tools provided by uncertainty quantification, we may easily describe the aggregate trends taking care of the structural heterogeneity of real traffic flows. 

Let us characterise the microscopic state of two interacting vehicles by means of their dimensionless and normalised speeds $v,\,v_\ast\in [0,\,1]$. We describe the post-interaction speeds $v^\prime,\,v_\ast^\prime$ in terms of the following scheme
\begin{equation}
\begin{cases}
	v^\prime=v+\gamma I(v,\,v_\ast;\,z)+D(v)\eta \\
	v_\ast^\prime = v_\ast. 
\end{cases}
\label{eq:micro_general}
\end{equation}
In~\eqref{eq:micro_general} the quantity $\gamma>0$ is a proportionality parameter whereas we indicated with $I$ a general interaction function depending on the pre-interaction states $v,\,v_\ast$ and on a set of uncertain quantities given by a random variable $z\in\R_+$, $z\sim\Psi$, where $\Psi:\R\to\R_+$ is a probability density function:
$$ \mathbb{P}(z\leq\bar{z})=\int_{-\infty}^{\bar{z}}\Psi(z)\,dz. $$
It is worth mentioning that~\eqref{eq:micro_general} is structurally anisotropic. Indeed, if a car behind another one modifies its speed this action does not induce that leading car to go faster or slower. Those assumptions are in agreement with follow-the-leader microscopic models, see~\cite{gazis1961} for the original microscopic modelling set-up and~\cite{Piccoli} for further investigation on their kinetic counterpart. 

The interaction describes the tendency to update the vehicle speed $v$ taking into account the speed of the other vehicle $v_\ast$ and an uncertain traffic composition. The term $D(v)\eta$ takes into account stochastic fluctuations due to possible deviations from the introduced deterministic behavioural scheme. In particular, $\eta \in \R$ is a centred random variable with finite non-zero variance, i.e.
$$ \langle\eta\rangle=0, \qquad \langle\eta^2\rangle=\sigma^2>0, $$
where $\langle\cdot\rangle$ denotes the expectation with respect to the law of the random variable $\eta$ and $\sigma$ is the standard deviation of $\eta$. The function $D:[0,1]\rightarrow \R$ expresses the local relevance of the stochastic fluctuations.

\subsection{The interaction function}
Recently, several interaction rules have been proposed in the literature on kinetic models for traffic flows in the absence of uncertainties. Generally, the interaction is modelled by considering separately the cases of acceleration, i.e. $v\le W$, or deceleration, i.e. $v>W$, where drivers decide their behaviour depending on a certain quantity $W$. If $W$ depends on the speed of the other vehicle, i.e. $W = g(v_\ast)$, then real binary interactions take place, $g$ being a given function acting as a threshold. See for instance~\cite{Guenter2002,Herty2018,Illner1999,tosin2018IFAC,Visconti2017} and also~\cite{Herty2010} for a review on possible interactions.

Here, taking inspiration from~\cite{tosin2018MMS}, we will consider instead the following interaction function:
\begin{equation}
	I(v,\,v_\ast;\,z) = P(\rho;\,z)(1-v) + (1-P(\rho;\,z))(P(\rho;\,z)v_\ast-v),
	\label{eq:I}
\end{equation}
where $P(\rho;\,z)\in [0,\,1]$ is the probability of acceleration. The interaction~\eqref{eq:I} is a convex combination between the tendency to travel with maximal speed, which is unitary in a dimensionless setting, and the necessity to adapt the speed to a fraction of the speed of the leading vehicle. Notice that~\eqref{eq:I} synthesises the post-interaction speed as a negotiation between acceleration and deceleration, without any threshold $W$ triggering sharply either of them.

The function $P(\rho;\,z)$ depends on the dimensionless traffic density $\rho\in [0,\,1]$ and on the uncertain quantity $z$. The form that we will consider in this chapter is the following:
\begin{equation}
	P(\rho;\,z)=(1-\rho)^z, \qquad z>0.
	\label{eq:defP}
\end{equation}
In particular, traffic flows with heterogeneous classes of vehicles are associated to different exponents of the function $P(\rho;\,z)$.

\begin{remark}
\label{rem:eta}
A compatibility condition for the interaction rule~\eqref{eq:micro_general} with the interaction function~\eqref{eq:I} is that the post-interaction speeds remain in the interval $[0,\,1]$. This can be guaranteed by imposing the following sufficient condition on the fluctuation $D(v)\eta$:
$$ \vert\eta\vert\leq c(1-\gamma), \qquad cD(v)\leq\min\{v,\,1-v\}, $$ 
where $c>0$ is an arbitrary constant, see~\cite{Tosin,tosin2018MMS}.
\end{remark}

\subsection{Kinetic description and equilibria}
Let $f=f(t,\,v;\,z)$ be the distribution function of the vehicles travelling with speed $v\in [0,\,1]$ at time $t\geq 0$ and belonging to the vehicle class $z\sim\Phi(z)$. Since microscopic interactions are binary and Remark~\ref{rem:eta} ensures that the post-interaction speeds remain always in $[0,\,1]$, we may rely on a Boltzmann-type equation for Maxwellian-like particles for the evolution of $f$, which in weak form is written as
\begin{equation}
	\frac{d}{dt}\int_0^1\varphi(v)f(t,\,v;\,z)\,dv=\frac{1}{2}\int_0^1\int_0^1\langle\varphi(v^\prime)-\varphi(v)\rangle f(t,\,v;\,z)f(t,\,v_\ast;\,z)\,dv\,dv_\ast,
	\label{eq:boltzmann_weak}
\end{equation}
where $\varphi:[0,\,1]\to\R$ is a test function. We refer the reader to~\cite{pareschi2013BOOK} for a detailed derivation of such a kinetic equation for collective phenomena.

From~\eqref{eq:boltzmann_weak} we may obtain information on the evolution of observable quantities, such as the mass of vehicles and their mean speed. In particular, letting $\varphi(v)=1$ we observe that the mass of the system is conserved since
$$ \frac{d}{dt}\int_0^1f(t,\,v;\,z)\,dv=0 $$
for all $z\in\R_+$. Therefore if at time $t=0$ the distribution $f$ is a probability density it remains so for all times $t>0$. Furthermore, for $\varphi(v) = v$ we have
\begin{equation*}
	\begin{split}
		\frac{d}{dt}V(t;\,z) &= \frac{\gamma}{2}\int_0^1\int_0^1 I(v,\,v_\ast;\,z)f(t,\,v;\,z)f(t,\,v_\ast;\,z)\,dv\,dv_\ast \\
		&= \frac{\gamma}{2}\left[P(\rho;\,z)(1-V)-(1-P(\rho;\,z))^2V\right],
	\end{split}
\end{equation*}
where $V(t;\,z)$ is the uncertain mean speed of the flow. For large times ($t\to +\infty$), we obtain its asymptotic profile which depends now uniquely on the system uncertainty and on the known traffic density $\rho$:
\begin{equation}
	V_\infty(\rho;\,z)=\frac{P(\rho;\,z)}{P(\rho;\,z)+(1-P(\rho;\,z))^2}.
	\label{eq:Vinf}
\end{equation}

Similarly, for the evolution of the energy we consider $\varphi(v) = v^2$ to obtain
\begin{equation*}
	\begin{split}
		\frac{dE}{dt} &= \frac{\gamma}{2}\int_0^1\int_0^1\left(\gamma I^2(v,\,v_\ast;\,z)+2vI(v,\,v_\ast;\,z)\right)f(t,\,v;\,z)f(t,\,v_\ast;\,z)\,dv\,dv_\ast \\
		&\phantom{=} +\frac{\sigma^2}{2}\int_0^1 D^2(v)f(t,\,v;\,z)\,dv \\
		&= \frac{\gamma^2P^2}{2}(1+E-2V)+\frac{\gamma^2(1-P)^2}{2}\left(\left(P^2+1\right)E-2PV^2\right) \\
		&\phantom{=} +\gamma P(V-E)+\gamma(1-P)(PV^2-E) \\
		&\phantom{=} +\frac{\sigma^2}{2}\int_0^1 D^2(v)f(t,\,v;\,z)\,dv.
	\end{split}
\end{equation*}
In the zero-diffusion limit $\sigma^2\to 0^+$ and for a traffic regime $\rho\in (0,\,1)$, the energy evolution reduces to 
\begin{equation*}
	\begin{split}
		\frac{dE}{dt} &= \frac{\gamma^2}{2}\Bigl[P^2(1+E-2V)+(1-P)^2\left(\left(P^2+1\right)E-2PV^2\right)\Bigr] \\
		&\phantom{=} +\gamma P(V-E)+\gamma(1-P)\left(PV^2-E\right).
	\end{split}
\end{equation*}
For large times, using~\eqref{eq:Vinf} we have
$$ E_\infty(\rho;\,z)=\left(\frac{P(\rho;\,z)}{P(\rho;\,z)+(1-P(\rho;\,z))^2}\right)^2=V_\infty^2(\rho;\,z). $$
Therefore, for all $\rho\in (0,\,1)$ the large time distribution is a Dirac delta $\delta(v-V_\infty(\rho;\,z))$ centred in the $z$-dependent asymptotic mean speed $V_\infty(\rho;\,z)$. It is interesting to observe that
in the traffic regime $\rho\to 0^+$, leading to $P\to 1$ (cf.~\eqref{eq:defP}), the evolution of the energy reduces to
$$ \frac{d}{dt}E(t;\,z)=\gamma\left(1-\frac{\gamma}{2}\right)(1-E(t;\,z)), $$
and therefore 
$$ E(t;\,z)=(E(0;\,z)-1)e^{-\gamma(1-\gamma/2)t}+1. $$
If $0<\gamma<2$ then $E(t;\,z)\to 1$ for every $z\in\R_+$. Since for $\rho\to 0^+$ we also have $V(t;\,z)\to 1^-$, the asymptotic distribution is again a Dirac delta $\delta(v-1)$, however independent of $z$. An analogous remark holds for $\rho\to 1^-$, for which $V(t;\,z)\to 0^+$ and the evolution of the energy is given by
$$ \frac{d}{dt}E(t;\,z)=-\gamma\left(1-\frac{\gamma}{2}\right)E(t;\,z), $$
leading now asymptotically to the Dirac delta $\delta(v)$ again independent of $z$.

A more detailed analysis of the aggregate behaviour of the kinetic model may be obtained looking at the equilibrium distribution for non-vanishing diffusion. Unfortunately, clear analytical insights are not easy to obtain in general, due to the complexity of the collision operator at the right-hand side of~\eqref{eq:boltzmann_weak}. In order to overcome this difficulty, we may however rely on the powerful asymptotic method of the quasi-invariant limit, see~\cite{cordier2005JSP,toscani2006CMS}. The idea is to consider the regime in which the parameters $\gamma$, $\sigma^2$ of the microscopic interactions are small, so that each interaction produces a small change of speed of the vehicles. At the same time, in order to balance the weakness of the interactions and to observe a trend in the limit $\gamma,\,\sigma^2\to 0^+$, one increases their rate by introducing the new time scale $\tau:=\gamma t/2$ and the scaled kinetic distribution function
$$ g(\tau,\,v;\,z):=f\left(\frac{2\tau}{\gamma},\,v;\,z\right), $$
which from~\eqref{eq:boltzmann_weak} is easily seen to solve the following Boltzmann-type equation:
$$ \frac{d}{d\tau}\int_0^1\varphi(v)g(\tau,\,v;\,z)\,dv=\frac{1}{\gamma}\int_0^1\int_0^1\langle\varphi(v^\prime)-\varphi(v)\rangle g(\tau,\,v;\,z)g(\tau,\,v_\ast;\,z)\,dv\,dv_\ast. $$
It is possible to prove, see~\cite{tosin2018MMS}, that if $\eta$ has moments bounded up to the third order then, in the limit $\gamma,\,\sigma^2\to 0^+$ with $\sigma^2/\gamma\to \lambda>0$, $g$ satisfies the following Fokker-Planck equation with non-constant coefficients:
\begin{equation}
	\partial_\tau g=\frac{\lambda}{2}\partial_v^2\left(D^2(v)g\right)-\partial_v\left(P(1+(1-P)U(\tau;\,z)-v)g\right),
	\label{eq:FP}
\end{equation}
where
$$ U(\tau;\,z):=\int_0^1vg(\tau,\,v;\,z)\,dv=V\left(\frac{2}{\gamma}\tau;\,z\right) $$
denotes the mean speed in the new time scale. Notice that $U(\tau;\,z)\to V_\infty(\rho;\,z)$ for $\tau\to +\infty$, because from the performed time scaling we infer that $\tau/t$ is constant for every $\gamma>0$.

We can now investigate the large time trends of equation~\eqref{eq:FP} more easily. The stationary distribution $g_\infty$ solves
$$ \frac{\lambda}{2}\partial_v(D^2(v)g_\infty(v;\,z))-(V_\infty(\rho;\,z)-v)g_\infty(v;\,z)=0, $$
whose general solution reads
\begin{equation}
	g_\infty(v;\,z)=C_{\lambda,\rho,z}\exp\left\{-\frac{2}{\lambda}\int\frac{V_\infty(\rho;\,z)-v}{D^2(v)}\,dv\right\},
	\label{eq:ginf}
\end{equation}
$C_{\lambda,\rho,z}>0$ being a normalisation constant such that $\int_0^1g_\infty(v;\,z)\,dv=1$ for all $z$. Depending on the choice of the function $D(v)$ different particular distributions may be obtained, a broad range of which has been investigated in~\cite{toscani2006CMS}.

The empirical speed distributions of traffic are typically supported in the bounded interval $[0,\,1]$, therefore classical probability densities, such as the normal and the log-normal ones, are not good approximations of the observable stationary profiles. It is worth remarking that the first attempts to fit speed profiles date back to the half of the past century, see~\cite{Berry1951}. In those original approaches, a deviation of the real data from the standard normal distribution was noticed, in particular, when the traffic density is close to the road capacity, for in that case the speed distribution becomes heavily skewed. More recently, beta distributions have been identified to fit quite well the experimental data of traffic speeds, see~\cite{maurya2016TRP,ni2018AMM} for a detailed account of statistical tests validating this conclusion. Interestingly, beta distributions may be obtained from~\eqref{eq:ginf} with the choice
$$ D(v)=\sqrt{v(1-v)}, $$
which produces
\begin{equation}
	g_\infty(v;\,z)=C_{\lambda,\rho,z}v^{\frac{2V_\infty(\rho;\,z)}{\lambda}-1}(1-v)^{\frac{2(1-V_\infty(\rho;\,z))}{\lambda}-1},
	\label{eq:ginfty}
\end{equation}
with
\begin{equation}
	C_{\lambda,\rho,z}:=\frac{1}{\operatorname{B}\!\left(\frac{2V_\infty(\rho;\,z)}{\lambda},\,\frac{2(1-V_\infty(\rho;\,z))}{\lambda} \right)},
	\label{eq:C}
\end{equation}
where $\operatorname{B}(\cdot,\,\cdot)$ is the beta function. Taking advantage of the known formulas for beta-distributed random variables, we easily see that, consistently with the kinetic model, the distribution~\eqref{eq:ginfty} has mean $V_\infty(\rho;\,z)$ and energy given by 
$$ E_\infty(\rho;\,z):=\int_0^1v^2g_\infty(v;\,z)dv=\frac{V_\infty(\rho;\,z)}{2+\lambda}\left(2V_\infty(\rho;\,z)+\lambda\right). $$

\subsection{Quantities of interest}
In order to validate our theoretical results by means of experimental data we need to define some quantities of interest to be observed. The advantage of our kinetic approach consists in an analytically closed and sufficiently rich description of the speed profiles emerging at equilibrium, which can be fruitfully compared with the information contained in the measured dataset.

Since the emerging equilibria are affected by the uncertainty brought by the parameter $z$, it is of paramount importance to define what we may observe if we compare theoretical profiles with experimental data. In view of the mixed traffic conditions, where different vehicles interact and modify their speeds, it is natural to measure expected quantities with respect to the $z$-uncertainty. Therefore, the reconstructed speed distribution has to be compared with the following expected distribution:
\begin{equation}
	\label{eq:QoI_g}
	\begin{split}
		\bar{g}_\infty(v) &:= \E_z\left[C_{\lambda,\rho,z}v^{\frac{2V_\infty(\rho;\,z)}{\lambda}-1}(1-v)^{\frac{2(1-V_\infty(\rho;\,z))}{\lambda}-1}\right] \\
		&= \int_{\R_+}C_{\lambda,\rho,z}v^{\frac{2V_\infty(\rho;\,z)}{\lambda}-1}(1-v)^{\frac{2(1-V_\infty(\rho;\,z))}{\lambda}-1}\Psi(z)\,dz,
	\end{split}
\end{equation}
where the normalization constant $C_{\lambda,\rho,z}$ has been defined in~\eqref{eq:C}. This poses the necessity to determine the more suited distribution $\Psi(z)$ that classifies the reaction strengths of the real flow.

Among the most studied diagrams for traffic dynamics, the fundamental diagram summarises macroscopic trends in terms of predicted flow in connection with the recorded density. The fundamental diagram may be obtained from the introduced kinetic modelling by looking at the equilibrium relationship between the traffic density and the $z$-averaged macroscopic flux of the vehicles, i.e. the mapping $\rho\mapsto\rho\bar{V}_\infty(\rho)$. Then the observable macroscopic trends are given by the following expected quantities:
\begin{equation}
	\label{eq:VE_bar}
	\begin{split}
		\bar{V}_\infty(\rho) &:= \E_z\left(V_\infty(\rho;\,z)\right)=\int_{\R_+}V_\infty(\rho;\,z)\Psi(z)\,dz, \\
		\bar{E}_\infty(\rho) &:= \E_z(E_\infty(\rho;\,z))=\int_{\R_+}E_\infty(\rho;\,z)\Psi(z)\,dz.
	\end{split}
\end{equation}
We may also recover the typical scattering observed in empirical fundamental diagrams by looking at the set
$$ \left\{(\rho,\,q)\in [0,\,1]\times\R_+\,:\,q\in\left[\rho\bar{V}_\infty(\rho)-\rho\varsigma(\rho),\,\rho\bar{V}_\infty(\rho)+\rho\varsigma(\rho)\right]\right\}, $$
where $\varsigma^2(\rho)$ is the $z$-variance of $V(\rho;\,z)$. Indeed, the superposition of different microscopic uncertainties due to different values of $z$ is able to explain the observable scattering in this type of diagrams. We refer the interested reader to~\cite{Tosin} for deeper insights into this approach and we mention also~\cite{Herty2018,puppo2016CMS,seibold2013NHM,Visconti2017} for alternative approaches.

\section{Description of traffic data}
\label{sec:3}
In this work we consider data published in~\cite{dataset}, which have been recently extracted from $15$ videos recorded by $5$ cameras in a single traffic direction on the German A3 motorway. The road section is composed by three lanes in each direction with a speed limit of $100~\si{km/h}$. The videos have been recorded in various traffic conditions, between 7:35 am and 8:00 am, for a total of $8305$ recorded vehicles. Each camera covers approximately $100~\si{m}$ of road, and they are spaced is such a way that the total recorded road length is $1~\si{km}$. Therefore, we may consider the collected data as representative of traffic dynamics in various congestion regimes.

The speeds of the vehicles are recovered out of the microscopic positions in consecutive frames. From time-labelled microscopic data, the evolution of macroscopic quantities characterising the flow can be computed, see~\cite{herty2018NHM,hogen}.

In order to recover the distributions of the microscopic speeds associated to a representative value of the density, we proceed as follows. For each single dataset, corresponding to one video recorded by one camera, we fix a sequence of $M+1$ equally spaced discrete times $\{t_k\}_{k=0}^M$, such that $t_{k+1}-t_k=:\delta t$, $t_0=0$ and $t_M=t_{\max}$, where $t_{\max}$ is the final observation time, in seconds, in the dataset (here, approximately $1500$ seconds for each video). Then, at each discrete time $t_k$ we count the number of vehicles $N(t_k)$ on the road and define the density as $\tilde{\rho}(t_k):=\frac{N(t_k)}{L}$, for $k=0,\,\dots,\,M$, where $L$ is the length of the section expressed in the unit length of $1~\si{km}$. Moreover, we collect all the microscopic speeds of the vehicles on the road at the corresponding discrete time $t_k$. We take $\delta t=1~\si{s}$ and apply this procedure separately to each camera, in order to avoid averaging between very different traffic conditions in different sections of the motorway.

All the computed values of the density $\tilde{\rho}$ are normalised with respect to the maximum allowed density on the road, i.e., the stagnation density $\rho_\text{max}$. Since this value is not represented well by the data, we prescribe it as a fixed constant, given by the ratio between the number of lanes and the typical vehicle length of $5~\si{m}$, plus $50\%$ of additional safety distance, so that
$$ \rho_\text{max}=\frac{3~\si{lanes}}{7.5~\si{m}} = 400~\si{vehicles/km}. $$
This approach allows us to define representative classes for the densities, identifying values in intervals of size $10^{-1}$. No density levels higher than $0.4$ have been observed in this dataset. However, as already noticed in~\cite{herty2018NHM}, this value is higher than the critical value of the density where a capacity drop in the flux is observed. The experimental distributions are obtained by considering all the microscopic speeds corresponding to a density value belonging to a given density level. The microscopic speeds are normalised with respect to the maximum detected speed in the whole data-set.

It is worth mentioning that the vehicles recorded in~\cite{dataset} have been automatically recognised through a 3D tracking system. Those vehicles can be classified in various classes, spanning from personal cars to bus and trucks with different loads. In particular, $29$ types were recognised to represent most of the vehicles in the videos. This natural observation is in agreement with what we introduced in Section~\ref{sec:2} and leads us to consider heterogeneous traffic conditions for each density levels.

\section{Calibration and results}
\label{sec:4}
In this section we show the effects of considering microscopic interactions with uncertainty and compare the extrapolated quantities of interest with reconstructions of real data~\cite{dataset}. In particular, we will focus on the comparison between experimental and theoretical speed distributions.

From data we may distinguish four density regimes $\rho\in\{0.1,\,0.2,\,0.3,\,0.4\}$. For each $\rho$, several approaches are possible when reconstructing distributions from microscopic quantities, here we opt for the \textit{kernel density estimation}. This technique considers a convolution of the empirical measures associated to the data with a smoothing kernel of given bandwidth $s>0$. Therefore, if $v_1,\,\dots,\,v_N$ are the microscopic speeds associated to the road density $\rho$, we consider the probability density function
$$ \hat{g}(v)=\frac{1}{Ns}\sum_{i=1}^NK\!\left(\frac{v-v_i}{s^2}\right), \qquad s^2>0, $$
being $K(x):=\frac{1}{\sqrt{2\pi}}e^{-\frac{x^2}{2}}$. Other possible approaches are the so-called weighted area rule~\cite{Hockney1981} and standard histograms. The kernel density estimation method may be regarded as a suitable mollification of the histograms.

Once the experimental speed distribution $\hat{g}(v)$ has been reconstructed, we need to estimate the proper uncertainty distribution $\Psi=\Psi(z)$ which makes the theoretical $\bar{g}_\infty(v)$, cf.~\eqref{eq:QoI_g}, as consistent as possible with $\hat{g}(v)$. Since mixed traffic conditions with different classes of vehicles are recorded, among the possible uncertainty distributions we may consider the case of a discrete random variable $z\in\{z_1,\,\dots,\,z_n\}\subset\R_+$ with law
\begin{equation}
	\label{eq:z_disc}
	\mathbb{P}(z=z_k)=\alpha_k\in [0,\,1], \qquad \sum_{k=1}^n\alpha_k=1,
\end{equation}
which leads to 
$$ \Psi(z):=\sum_{k=1}^n\alpha_k\delta(z-z_k), $$
where $\delta(z-z_k)$ is the Dirac delta distribution centred in $z=z_k$. In this case, we have that~\eqref{eq:QoI_g} is 
\begin{equation}
	\label{eq:bar_g_disc}
	\bar{g}_\infty(v)=\sum_{k=1}^n\alpha_kC_{\lambda,\rho,z_k}v^{\frac{2V_\infty(\rho;\,z_k)}{\lambda}}(1-v)^{\frac{2(1-V_\infty(\rho;\,z_k))}{\lambda}-1}.
\end{equation}
Therefore, in general we obtain observable speed distributions depending on $2n$ parameters, specifically $z_1,\,\dots,\,z_n$, $\alpha_1,\,\dots,\,\alpha_{n-1}$ and $\lambda$, since $\alpha_n=1-\sum_{k=1}^{n-1}\alpha_k$.

In order to compare~\eqref{eq:bar_g_disc} with the experimental $\hat{g}$ obtained through the kernel density estimation we solve the following constrained optimisation problem:
\begin{equation}
	\min_{(z_1,\,\dots,z_n,\,\alpha_1,\,\dots,\,\alpha_{n-1},\,\lambda)}\mathcal{J}(\hat{g},\,\bar{g}_\infty),
	\label{eq:min}
\end{equation}
where $\mathcal{J}$ is the cost functional
\begin{equation}
	\mathcal{J}(\hat{g},\,\bar{g}_\infty)=\left(\int_0^1\left\vert\hat{g}(v)-\bar{g}_{\infty}(v)\right\vert^2\,dv\right)^{1/2},
	\label{eq:J}
\end{equation}
namely the $L^2$ norm of the difference between $\hat{g}$ and $\bar{g}_\infty$. Problem~\eqref{eq:min} has to be solved under the constraints
\begin{align*}
	z_k\geq 0 & & \forall\,k=1,\,\dots,\,n-1 \\
	0\leq\alpha_k\leq 1 & & \forall\,k=1,\,\dots,\,n-1 \\
	\lambda\geq 0.
\end{align*}
This optimisation procedure has been performed through the standard \texttt{fmincon} algorithm of \textsc{Matlab}\textsuperscript{\textregistered}.
 
\setlength{\tabcolsep}{12pt}
\begin{table}[!t]
\begin{center}
\caption{Values of the estimated set of parameters $(z_1,\,z_2,\,\alpha_1,\,\lambda)$ obtained through the optimisation procedure~\eqref{eq:min} relative to four density levels. In the last column we report the value of  the cost functional $\mathcal{J}$~\eqref{eq:J}.}
\label{tab:parameters}
\begin{tabular}{cccccc}
\hline
$\rho$ & \multicolumn{4}{c}{Parameters} & Cost \\
\hline
& $z_1$ & $z_2$ & $\alpha_1$ & $\lambda$ & $\mathcal{J}(\hat{g},\,\bar{g}_\infty)$ \\ 
\hline\hline
\multirow{1}{*}{0.1}	& 8.365 & 8.365 & 0.500 & 0.1185 & 0.2533 \\ 
\hline
\multirow{1}{*}{0.2} & 6.475 & 4.140 & 0.256 & 0.1185 & 0.2874 \\
\hline
\multirow{1}{*}{0.3}	& 4.411 & 2.741 & 0.528 & 0.0806 & 0.4954 \\
\hline
\multirow{1}{*}{0.4} & 3.186 & 2.073 & 0.425 & 0.0860 & 0.7603 \\
\hline
\end{tabular}
\end{center}
\end{table}

In Table~\ref{tab:parameters} we summarise the parameters obtained in the case $n=2$, where interactions are characterised by two possible strengths corresponding to the simplified case where two classes of vehicles are considered. From~\eqref{eq:z_disc} we observe that $\alpha_1+\alpha_2=1$. Since the iterative algorithm used for solving the optimisation problem is sensitive to the initial guesses $z_1^0,\,\dots,\,z_N^0,\,\alpha_1^0,\,\dots,\,\alpha_N^0,\,\lambda^0$, we have performed a further analysis by solving different optimisation problems spanning uniformly distributed values of $z_1^0,\,z_2^0,\,\lambda^0\in [0.1,\,3]$ ($5$ values), $\alpha_1^0\in [0.1,\,0.9]$ ($3$ values). From this analysis we have obtained $375$ sets of parameters. We have selected the optimal set through the minimisation of the $L^1$ error between the empirical and the expected mean speed and energy, i.e.
$$ \operatorname{Err}_1=\left\vert\int_0^1v\hat{g}(v)\,dv-\bar{V}_\infty(\rho)\right\vert, \qquad
	\operatorname{Err}_2=\left\vert\int_0^1v^2\hat{g}(v)\,dv-\bar{E}_\infty(\rho)\right\vert, $$
where $\bar{V}_\infty$, $\bar{E}_\infty$ have been defined in~\eqref{eq:VE_bar}.

\begin{figure}[!t]
\centering
\includegraphics[scale=0.48]{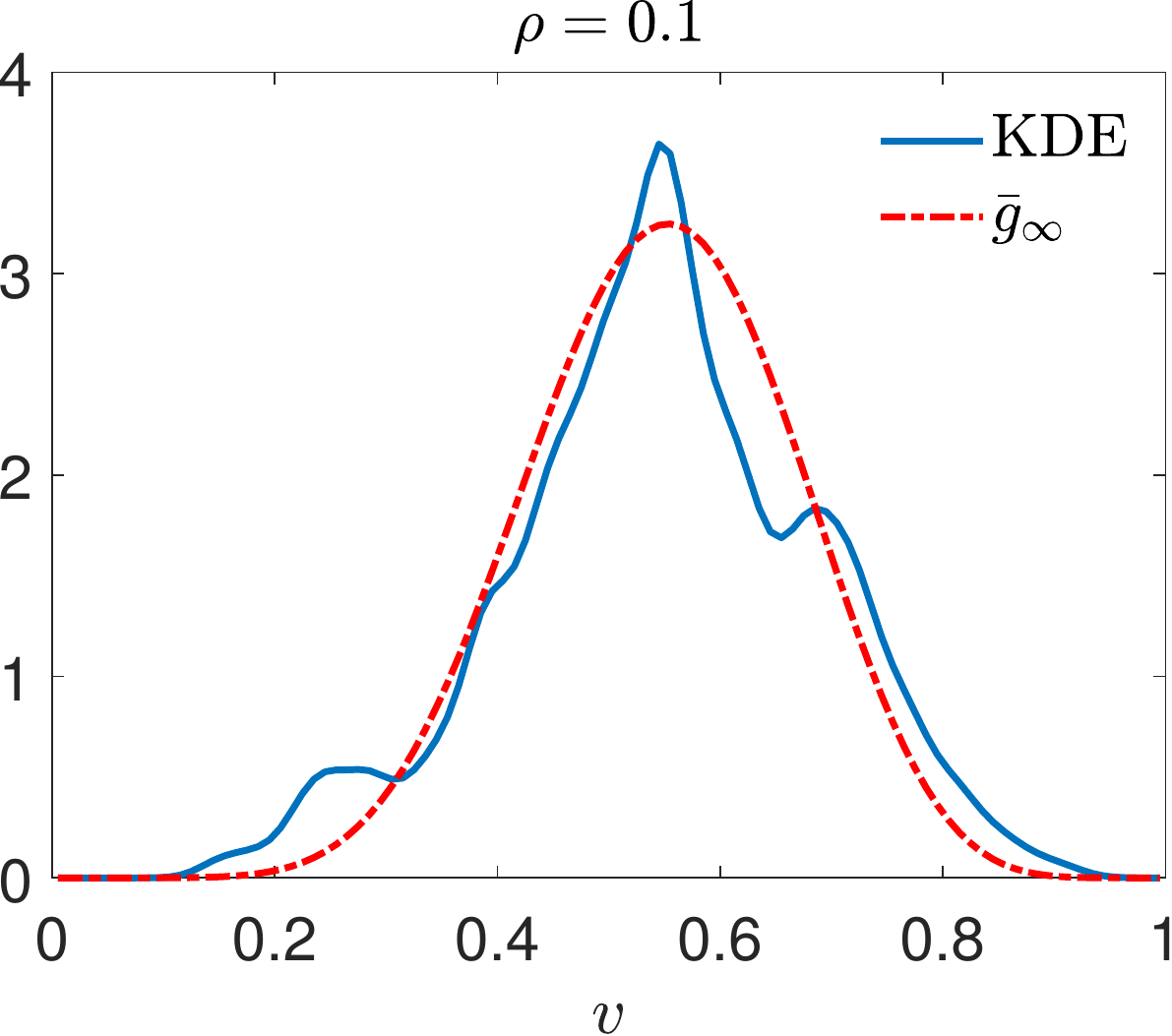}
\includegraphics[scale=0.48]{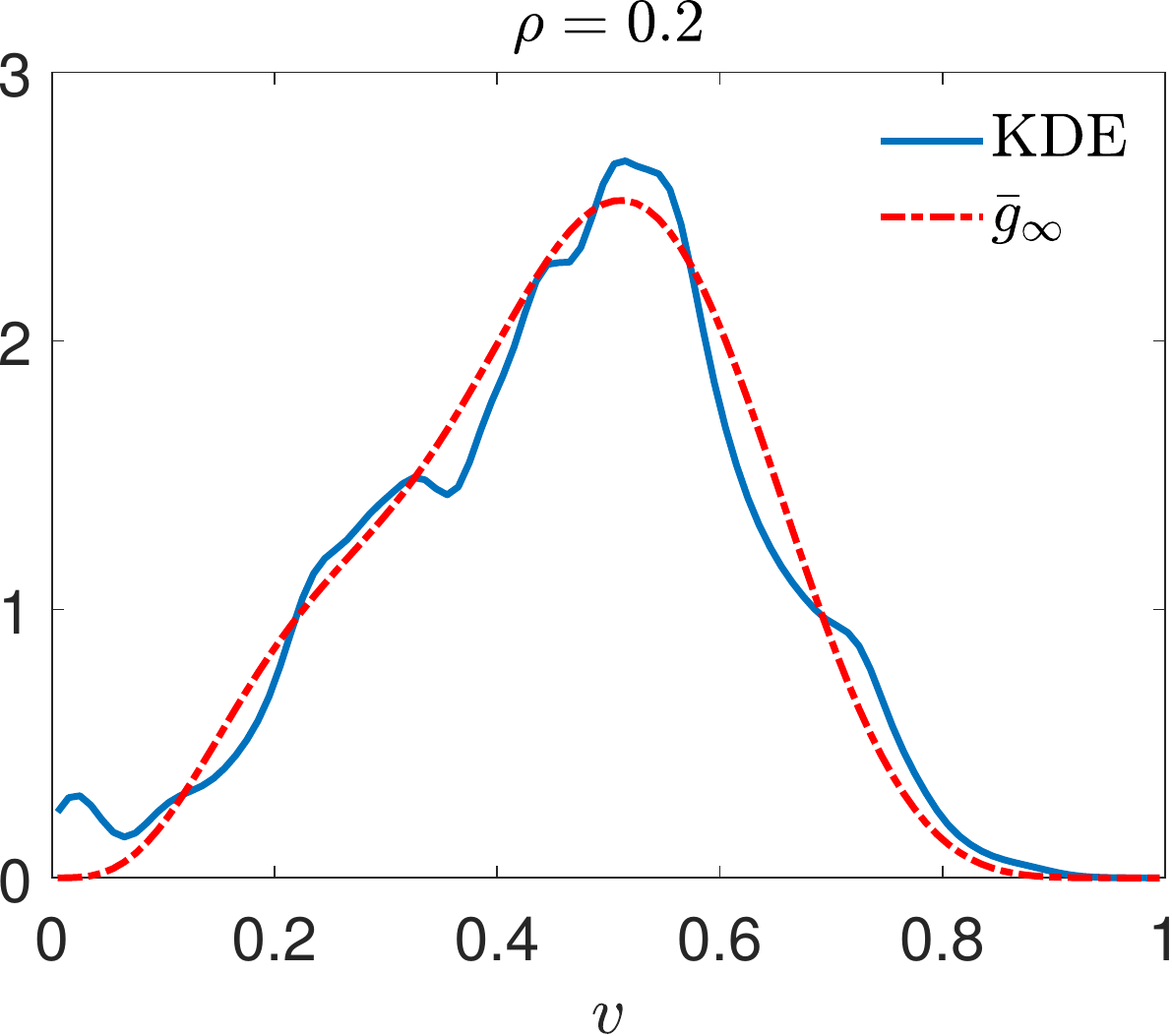} \\
\includegraphics[scale=0.48]{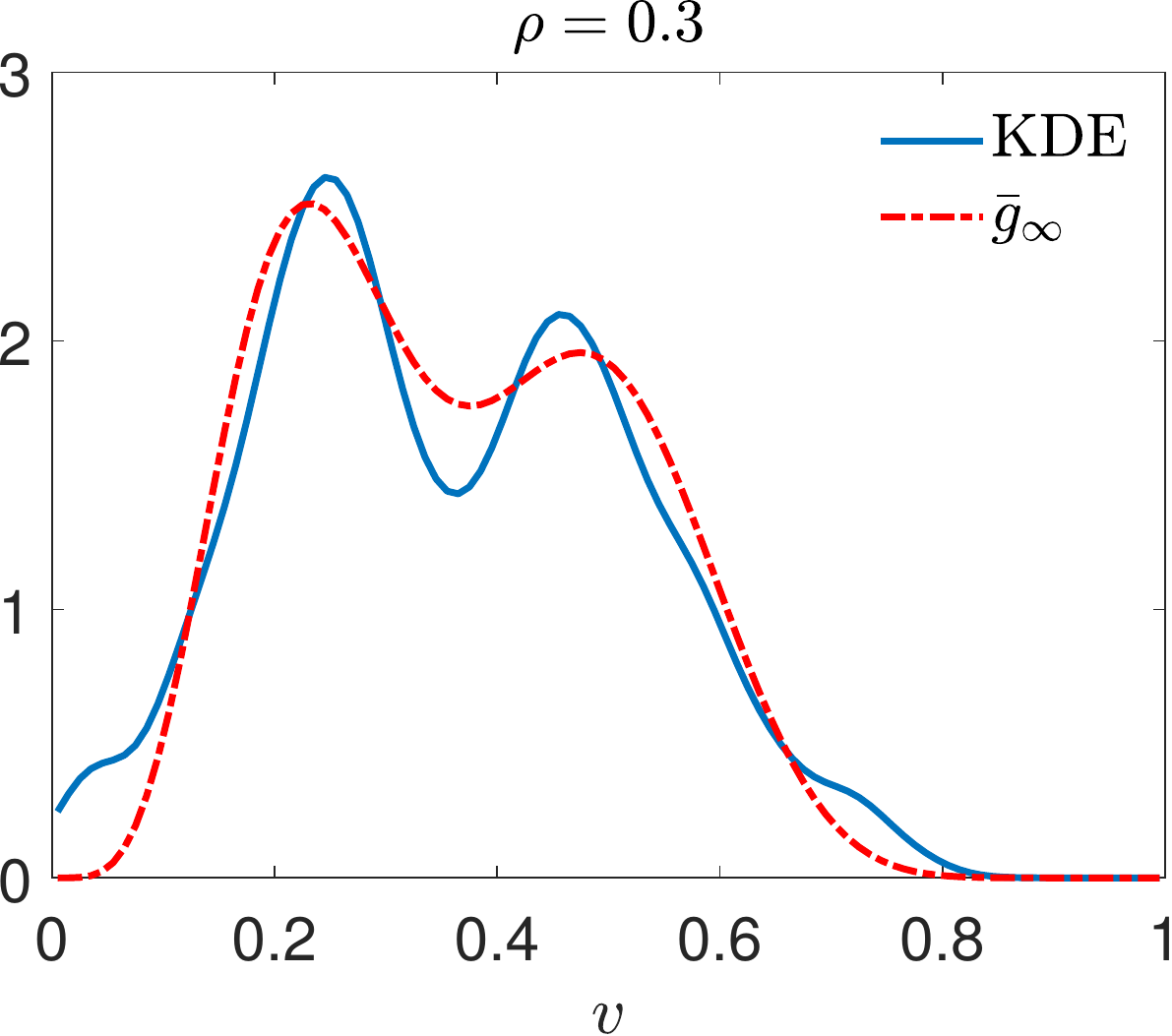}
\includegraphics[scale=0.48]{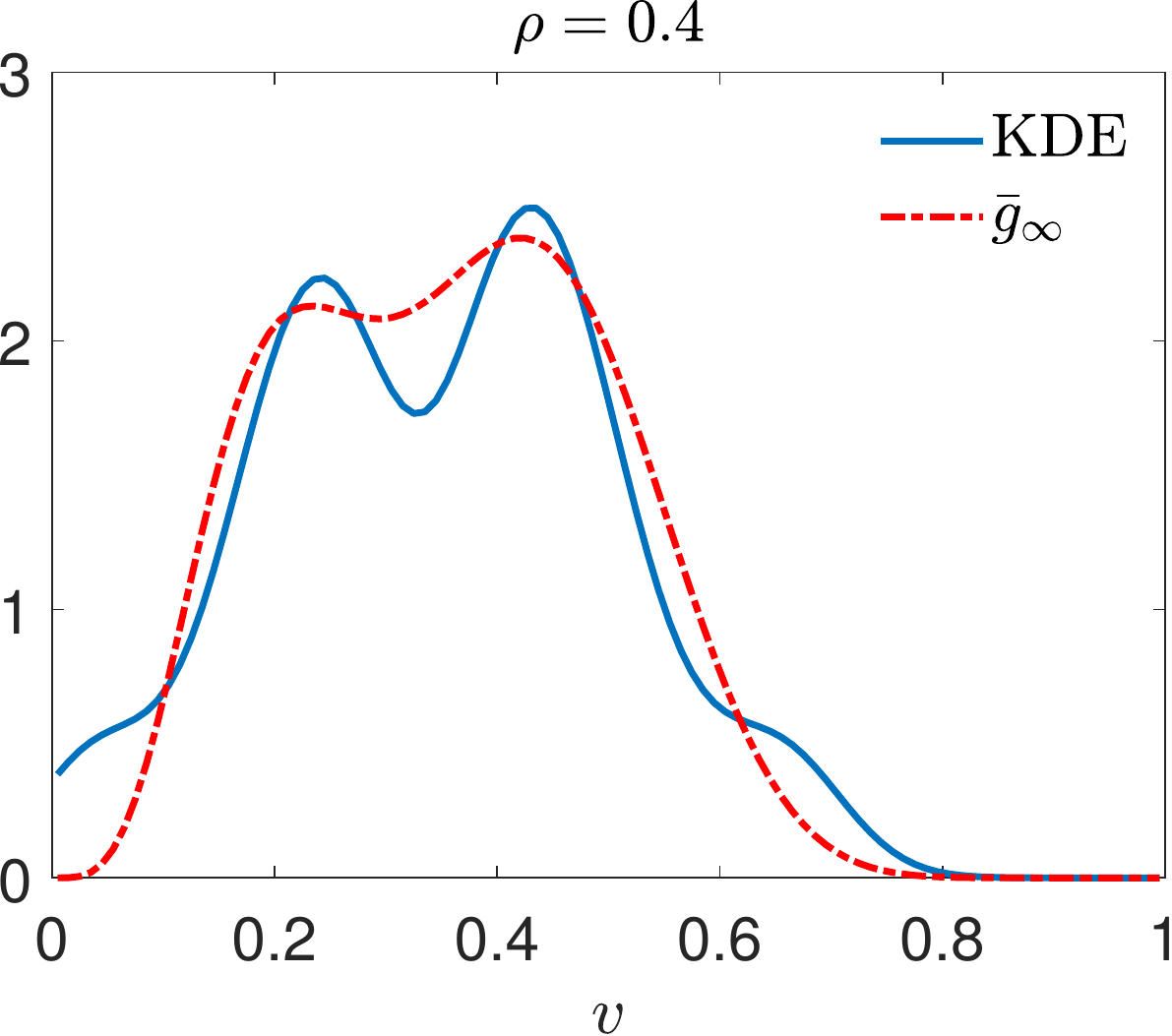}
\caption{Comparison between the densities reconstructed from rough data through the kernel density estimation and $\bar{g}_\infty$ defined by the optimal set of parameters reported in Table~\ref{tab:parameters}.}
\label{fig:1}
\end{figure}
 
In Figure~\ref{fig:1} we plot the reconstructed distributions in the density regimes $\rho\in\{0.1,\,0.2,\,0.3,\,0.4\}$ together with the $\bar{g}_\infty$ obtained with the optimal parameters in the case of uncertainty of the form~\eqref{eq:z_disc} and $n=2$. Remarkably, in free traffic regimes, i.e. for $\rho=0.1,\,0.2$, the single peak appearing in $\hat{g}$ is nicely captured by $\E[g_\infty]$. Furthermore, in congested traffic regimes, i.e. for $\rho\in\{0.3,\,0.4\}$, $\hat{g}$ shows a bimodal trend which is in turn nicely captured by $\E[g_\infty]$.

\setlength{\tabcolsep}{12pt}
\begin{table}[!t]
\begin{center}
\caption{Values of the extrapolated first and second order moments of the distributions $\hat{g}$ and $\bar{g}_\infty$, the latter being the optimised one with the parameters given in Table~\ref{tab:parameters}.}
\label{tab:moments}
\begin{tabular}{cccccc}
\hline
$\rho$ & \multicolumn{4}{c}{Extrapolated moments} \\
\hline
& $\hat{V}$ & $\bar{V}_\infty$ & $\hat{E}$ & $\bar{E}_\infty$ \\ 
\hline\hline
\multirow{1}{*}{0.1}	& 0.6009 & 0.6016 & 0.3498 & 0.3475 \\ 
\hline
\multirow{1}{*}{0.2} & 0.5132 & 0.5132 & 0.2673 & 0.2610 \\
\hline
\multirow{1}{*}{0.3} & 0.4010 & 0.3996 & 0.1746 & 0.1706 \\
\hline
\multirow{1}{*}{0.4} & 0.3975 & 0.3929 & 0.1691 & 0.1626 \\
\hline
\end{tabular}
\end{center}
\end{table}

To further validate the proposed approach, in Table~\ref{tab:moments} we report the values of the first and second moments, namely the mean speed and the energy, extrapolated from the empirical and theoretical distributions $\hat{g}$ and $\bar{g}_\infty$, respectively. In particular, $\bar{V}_\infty$ and $\bar{E}_\infty$ defined in~\eqref{eq:VE_bar} are compared with
$$ \hat{V}=\int_0^1v\hat{g}(v)\,dv, \qquad \hat{E}=\int_0^1v^2\hat{g}(v)\,dv. $$
Interestingly, we observe that the comparison gives perfectly consistent results in each density regime.

\section*{Conclusions}

In this work we have proposed an approach for the effective reconstruction of traffic speed distributions starting from the assessment of microscopic rough data. The techniques here developed are rooted in the statistical framework of the Boltzmann-type kinetic theory for multi-agent systems~\cite{pareschi2013BOOK}. In particular, the microscopic interactions among the vehicles are assumed to be binary and are defined starting from basic assumptions on the driver behaviour consistent with follow-the-leader-type dynamics. The additional formalism of the uncertainty quantification has allowed us to consider real traffic flows, in which mixed conditions are often observed due e.g., to the simultaneous presence of different types of vehicles. We have translated this feature in uncertain microscopic interactions~\cite{tosin2018CMS}, the uncertain parameter $z$ being one which affects the reaction strength of the vehicles in acceleration/deceleration.

From our kinetic approach, in particular in the asymptotic regime of the quasi-invariant interactions~\cite{toscani2006CMS}, we have been able to compute analytical equilibrium speed distributions, which fit nicely the empirically interpolated beta distributions~\cite{maurya2016TRP,ni2018AMM} and maintain also an explicit dependence on the uncertainty parameter $z$. By estimating the statistical distribution of $z$ via an optimisation procedure grounded on real traffic data recorded on the German A3 motorway~\cite{dataset}, we have recovered also more complex speed distributions, such as e.g., bimodal ones emerging in medium density traffic regimes, as the $z$-averaged superposition of ``elementary'' beta distributions.

We believe that these results pave the way to a physically sound and mathematically consistent procedure for the reconstruction and quantification of microscopic uncertainties naturally present in collective phenomena, which may have a considerable impact at larger scales.

\section*{Acknowledgements}
This research was partially supported by the Italian Ministry of Education, University and Research (MIUR) through the ``Dipartimenti di Eccellenza'' Programme (2018-2022) -- Department of Mathematical Sciences ``G. L. Lagrange'', Politecnico di Torino (CUP:E11G18000350001) and Department of Mathematics ``F. Casorati'', University of Pavia; and through the PRIN 2017 project (No. 2017KKJP4X) ``Innovative numerical methods for evolutionary partial differential equations and applications''.

This work is also part of the activities of the Starting Grant ``Attracting Excellent Professors'' funded by ``Compagnia di San Paolo'' (Torino) and promoted by Politecnico di Torino.

A.T. and M.Z. are members of GNFM (Gruppo Nazionale per la Fisica Matematica) of INdAM (Istituto Nazionale di Alta Matematica), Italy.

The research of M.H. and G.V. is funded by the Deutsche Forschungsgemeinschaft (DFG, German Research Foundation) under Germany's Excellence Strategy -- EXC-2023 Internet of Production -- 390621612.

M.H. and G.V. acknowledge the ISAC institute at RWTH Aachen, Prof. M. Oeser, Dr. A. Fazekas, MSc. M. Berghaus and MSc. E. Kall\'{o} for kindly providing the trajectory data within the DFG project ``Basic Evaluation for Simulation-Based Crash-Risk-Models: Multi-Scale Modelling Using Dynamic Traffic Flow States''.

\bibliographystyle{plain}

%\bibliography{HmTaVgZm-speed_reconstruction}
\end{document}